%% file: DPF2019_template.tex
\documentclass[11pt]{article}
\usepackage[margin=1in]{geometry}
\usepackage{graphicx}
\usepackage{float}
\usepackage{caption}
\usepackage{subcaption}

\input econfmacros.tex 

\def\Title#1{\begin{center} {\Large {\bf #1} } \end{center}}
\def\Author#1{\begin{center} {\normalsize {\sc #1} } \end{center}}
\def\Institution#1{\begin{center} {\normalsize {\it #1} } \end{center}}
\def\Abstract#1{\noindent {\normalsize {\bf Abstract:} {\normalfont #1}}}
\def\Conference{\vspace{4mm}\begin{raggedright} {\normalsize {\it Talk presented at the 2019 Meeting of the Division of Particles and Fields of the American Physical Society (DPF2019), July 29--August 2, 2019, Northeastern University, Boston, C1907293.} } \end{raggedright}\vspace{4mm}}

\begin{document}

%
%

\Title{Cherenkov Light in Liquid Scintillator at the NOvA Experiment}

\Author{Shiqi Yu}

\Institution{Argonne National Laboratory\\ Illinois Institue of Technology}

\Abstract{NOvA is a long-baseline neutrino oscillation experiment with two functionally identical liquid scintillator tracking detectors, i.e. the Near Detector (ND) and the Far Detector (FD). One of NOvA's physics goal is to measure neutrino oscillation parameters by studying $\nu_e$ appearance and $\nu_{\mu}$ disappearance with the Neutrinos at the Main Injector (NuMI) beam at Fermi National Accelerator Laboratory. An accurate light model is a prerequisite for precise charged particle energy estimation in the detectors. Particle energy is needed in event reconstruction and classification, both of which are critical to constraining neutrino oscillation parameters. In this paper, I will explain the details of the data-driven tuning of the Cherenkov model and the impact of the new scintillator model.}

\Conference

\section{Introduction}

The NOvA detector simulation describes the scintillation light produced in the liquid scintillator and captured by the NOvA photon detectors. The scintillation light produced by charged particles shows characteristics that depend on the incident particle type, the energy of the particle and the properties of the scintillator. 

Two major effects to be understood in the NOvA detector light simulation are molecular scintillation light production with its quenching effect, and Cherenkov light production. These light components are tuned in MC simulation with use of cosmic ray muon data and ND data coming from NuMI beam.
 
\section{NOvA Light Model}
NOvA employs Birks' Law (Eq.~\ref{birks}) to describe the quenching effect. The decrease of fluorescence intensity of a given substance is described by the light yield per path length:

\begin{equation}
\label{birks}
\frac{dS}{dr} = \frac{A\frac{dE}{dr}}{1+k_B\frac{dE}{dr}},
\end{equation}

\noindent where $S$ represents the scintillation response, $r$ represents residual range, defined as the distance from the current position to the end of the track, $A$ is the scintillator efficiency, and $k_B$ is the quenching, or Birks, constant which depends on the scintillator material. NOvA employs the Birks constant that has most recently been evaluated by the NOvA collaboration~\cite{Dubnapaper}. 

Cherenkov radiation light is emitted when a charged particle passes through the scintillator at a speed greater than the velocity of light in the medium. In NOvA, the final light model includes the scintillation light yield, described by the Birks model, and the Cherenkov light production. After the final light model is applied to the MC simulation, it is expected that the MC simulation will accurately describe the data. By using Eq.~\ref{lightlevel}, the proportion of photons contributed by the Birks model and the Cherenkov light production is parameterized so that the total number of photons in the MC simulation matches the data observation:
\begin{equation}\label{lightlevel}
N_\gamma = F_{view}(Y_sE_{Birks} + \epsilon_C C_{\gamma}).
\end{equation}

\noindent Here, $N_\gamma$ is the total number of photons produced in the scintillator before being collected by the fiber, $Y_s$ is the scintillator brightness per unit energy deposition, $E_{Birks}$ represents the energy deposition simulated by the Birks model given above (Eq.~\ref{birks}), $\epsilon_C$ is the scintillator efficiency for Cherenkov photons, and $C_{\gamma}$ is the number of Cherenkov photons produced by the charged particle based on MC simulation. These model parameters ($Y_s$, $E_{Birks}$, $\epsilon_C$ and $C_{\gamma}$) are independent of views or detectors. $F_{view}$ is a scaling factor for the $xz$ ($yz$) view of the ND (FD) detector in order to characterize the differences between the views and detectors. A least-squares fit between data and MC events is performed to extract the parameters.

retune the light level model by incorporating a few measurements of the Birks constant performed by the Dubna group into NOvA detector simulation. 

In NOvA light model tuning, the ratio between Cherenkov and Birks effect is tuned incorporating a few measurements of the Birks constant ~\cite{Dubnapaper} into NOvA detector simulation. NOvA light model has been one of the significant systematic uncertainties in NOvA's oscillation analysis. 

\section{NOvA Light Model Tuning}
In light model tuning, four different sets of data and MC samples are used to perform a joint fit on the light level parameters. These four sets of samples are: ND cosmic rays (ND CRY), FD cosmic rays (FD CRY), muon and proton tracks of $\nu_{\mu}$ interactions in the ND from the NuMI beam. Four different sets of selections are applied to get the four samples which are used in the tuning. 

For cosmic sample, the selections based on detector geometry are applied so that the starting point of a track should be outside of the detector volume with a distance to the walls of detectors.

Similarly, a set of selections is applied to ND beam data and MC files to select out the quasi-elastic-like (QE-like) $\nu_{\mu}$ events in the ND and further separate the tracks of the selected $\nu_{\mu}$ events into muon track sample and proton track sample. Since muon and proton tracks have different behaviors in providing the Cherenkov photons, both of these two samples are of interest to constrain the Cherenkov photon contributions in NOvA light model.

Two dimensional spectrum is used for fitting the light model in Eq.~\ref{lightlevel}. It is constructed in the following way: x-axis is distance of a hit to the end of the track, so called residual track length, and y-axis is photon-electron pre-calibrated energy (pe) per unit path length of the hit. 

With light model tuning, an improvement on data and MC agreement on distribution of pe per unit path length is expected, so that the following calibration can take the corrected photon-electron raw energy as input and further calibrate hit energy into physics unit (GeV). 

\begin{figure}[H]\centering
  \begin{subfigure}{0.48\textwidth}
    \includegraphics[width=0.48\textwidth]{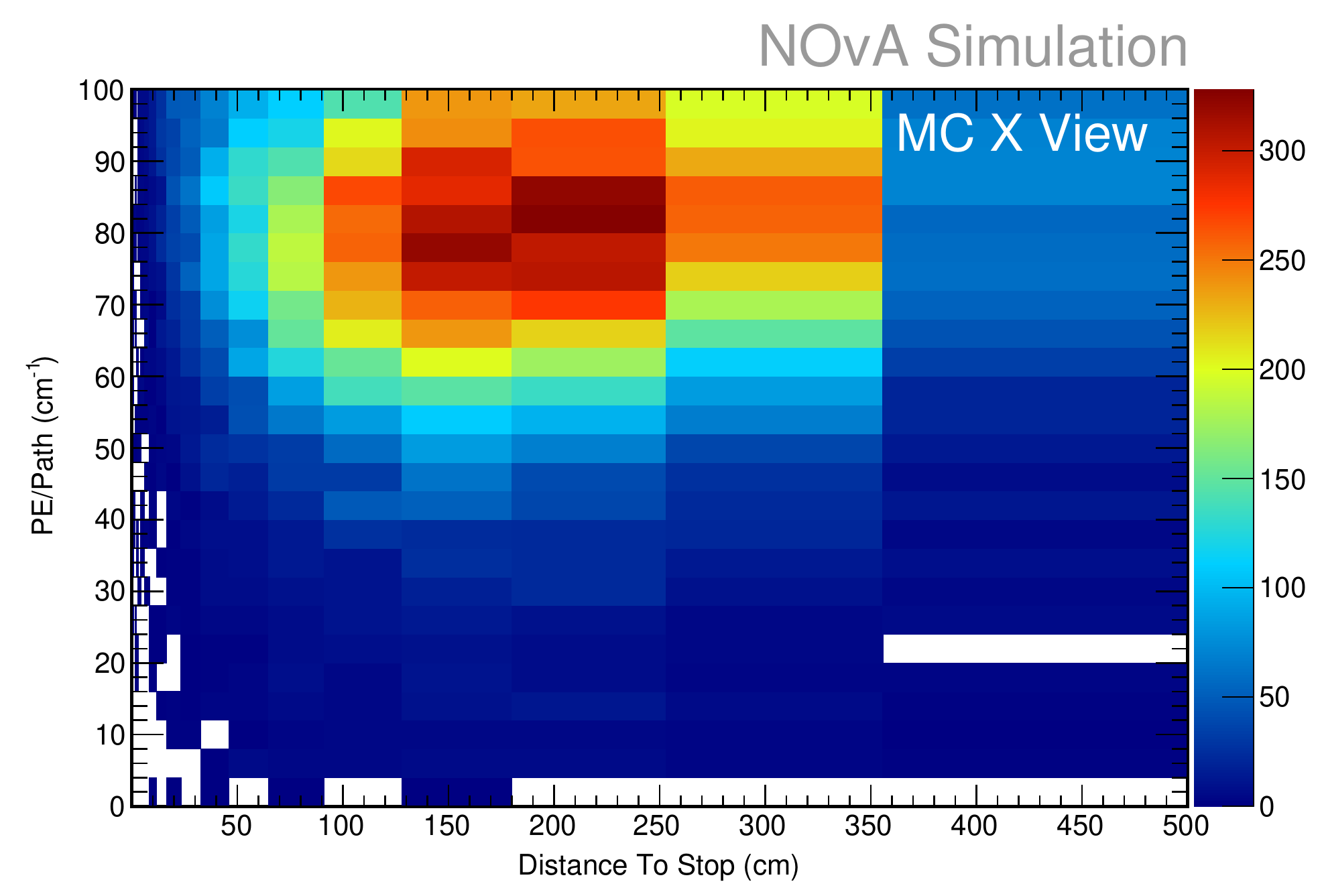}
    \includegraphics[width=0.48\textwidth]{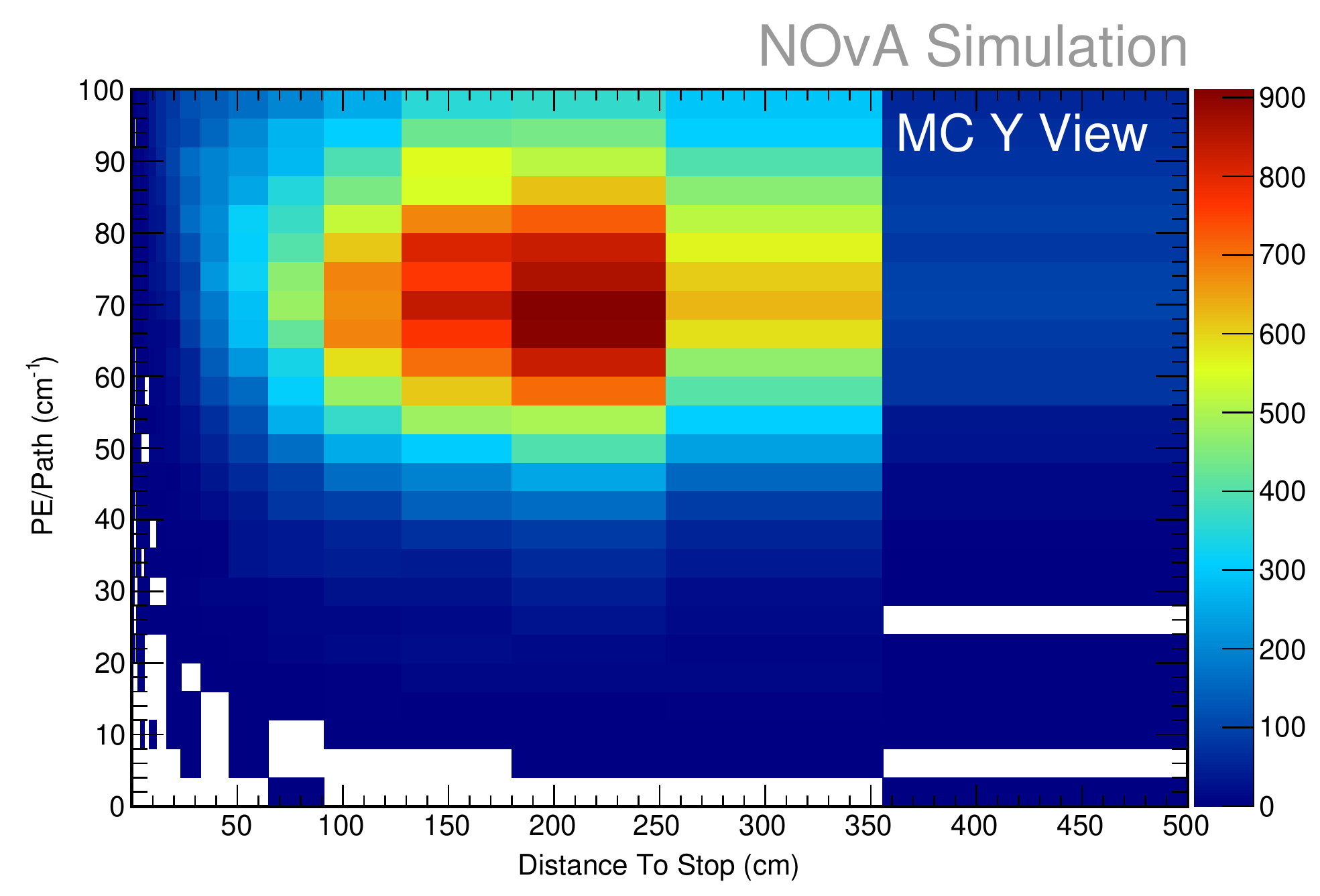}\\
    \includegraphics[width=0.48\textwidth]{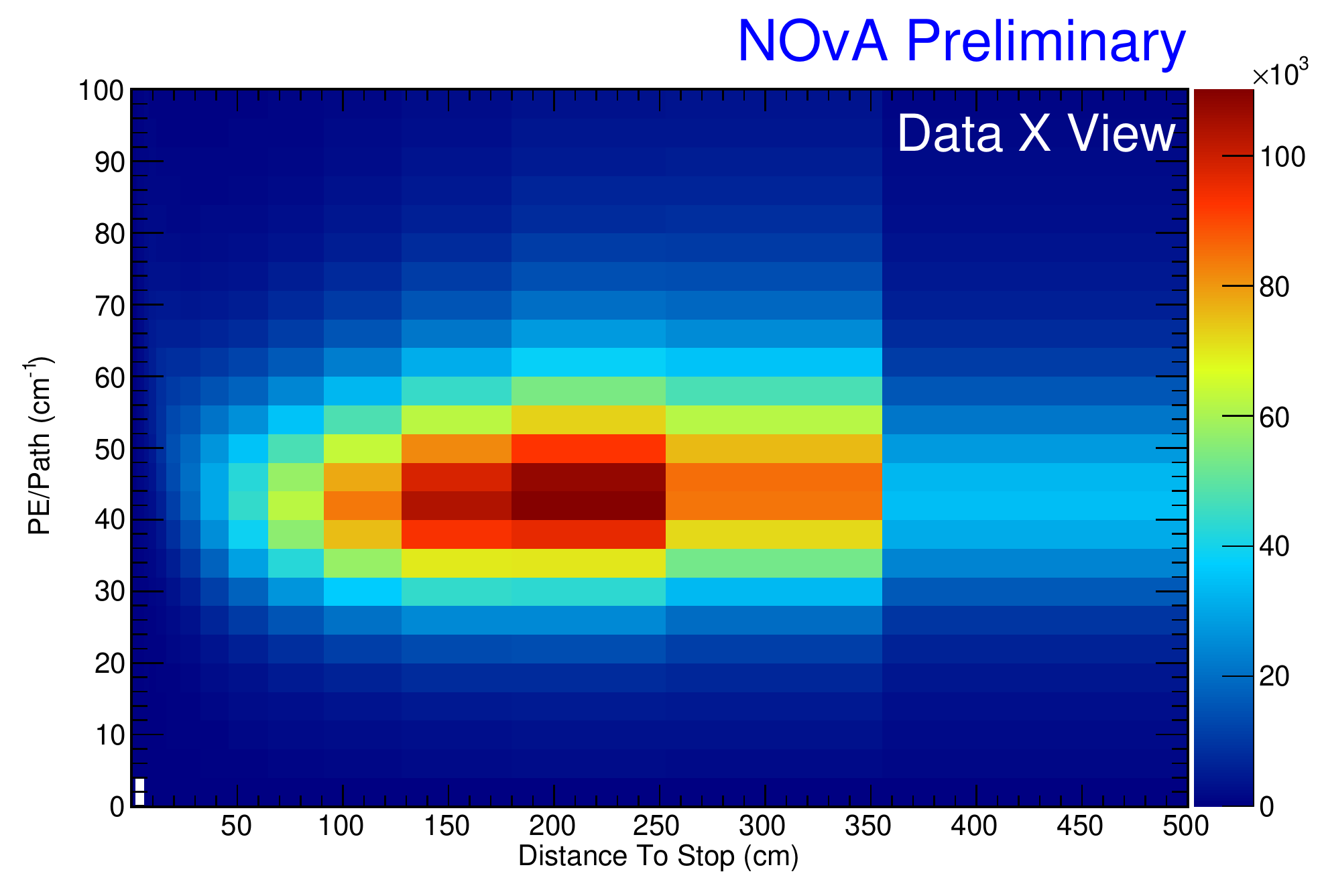}
    \includegraphics[width=0.48\textwidth]{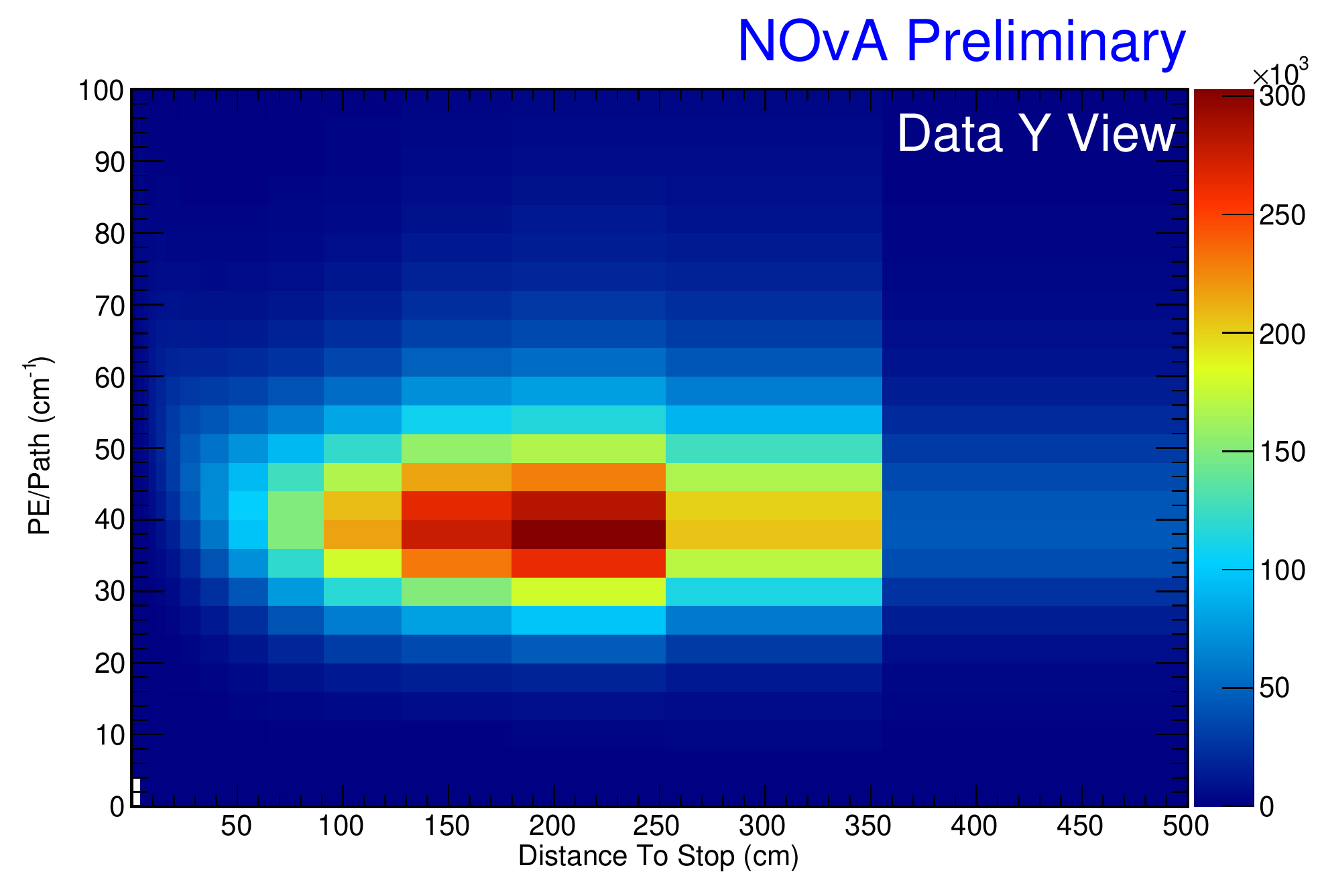}  
    \caption{Selected ND cosmic sample.}
  \end{subfigure}
  \begin{subfigure}{0.48\textwidth}
    \includegraphics[width=0.48\textwidth]{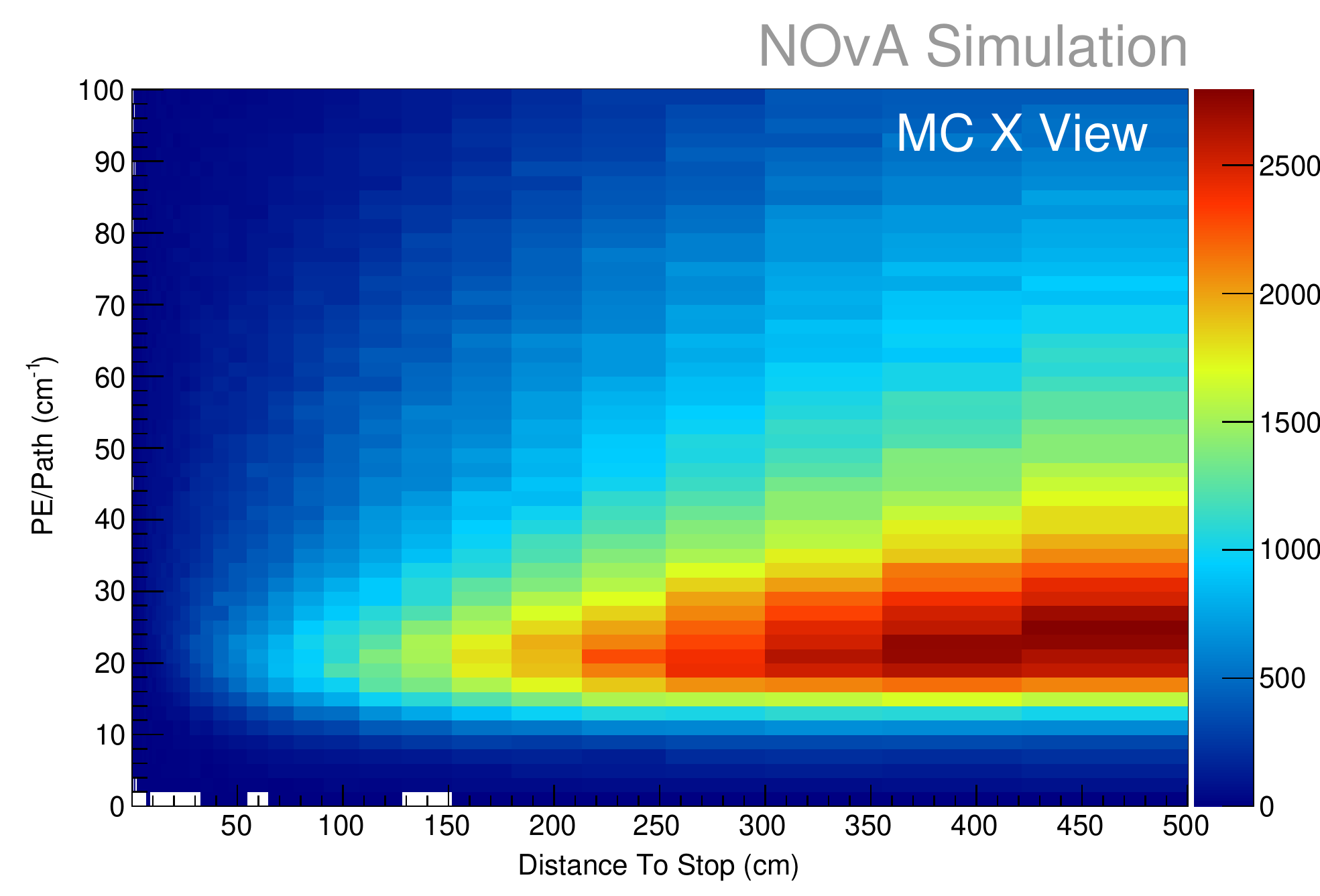}
    \includegraphics[width=0.48\textwidth]{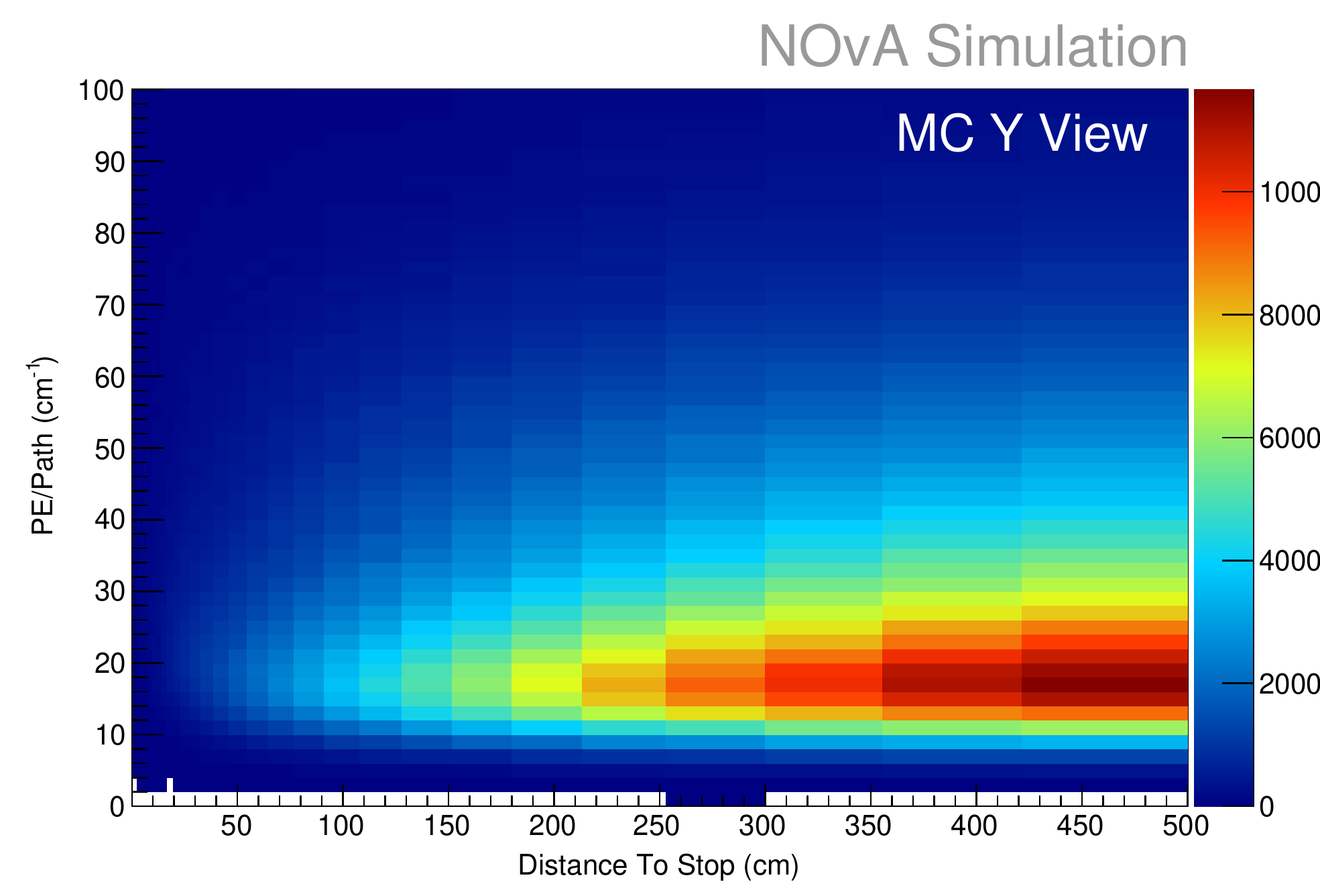}\\
    \includegraphics[width=0.48\textwidth]{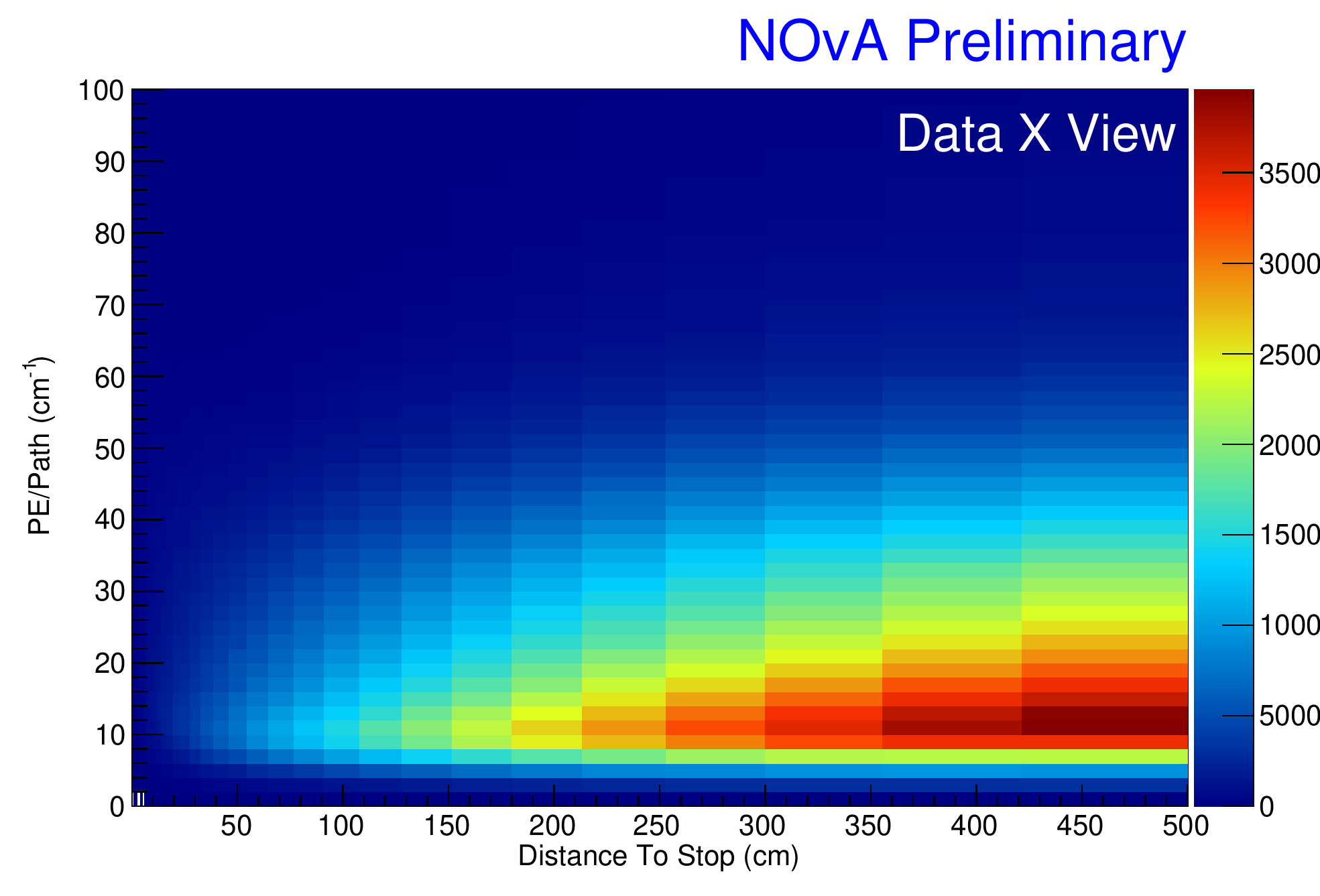}
    \includegraphics[width=0.48\textwidth]{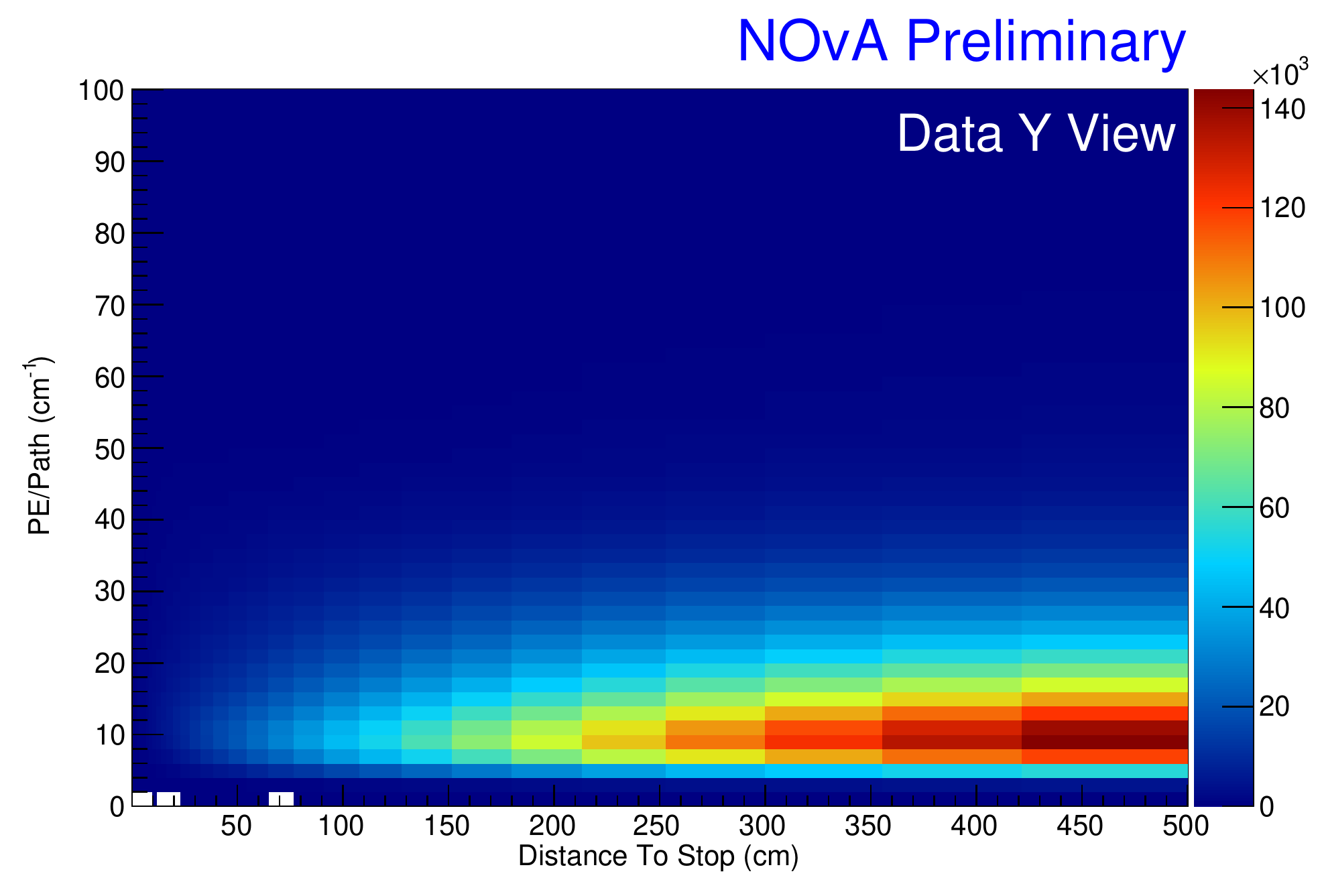}
    \caption{Selected FD cosmic sample.}
  \end{subfigure}

  \begin{subfigure}{0.48\textwidth}
    \includegraphics[width=0.48\textwidth]{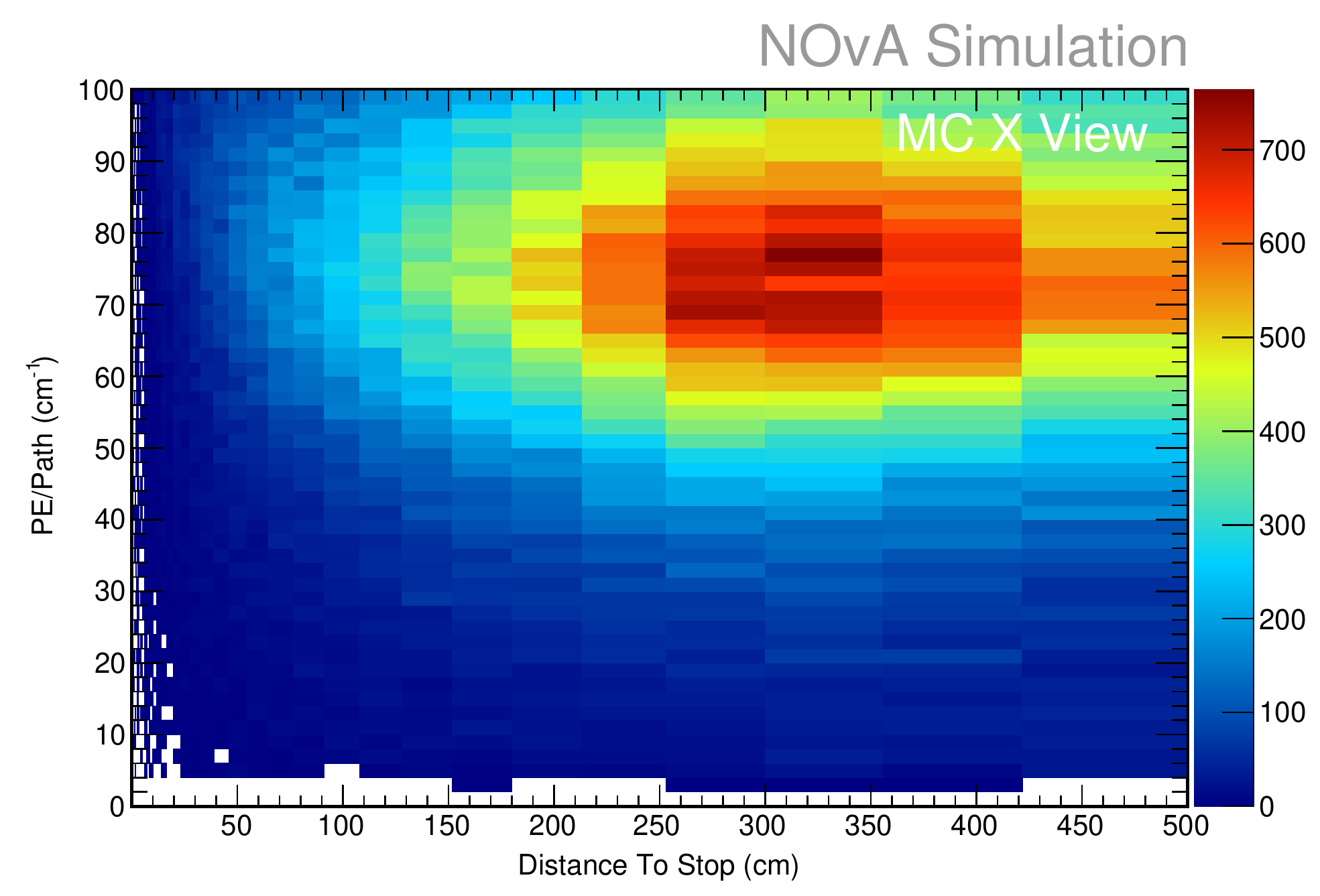}
    \includegraphics[width=0.48\textwidth]{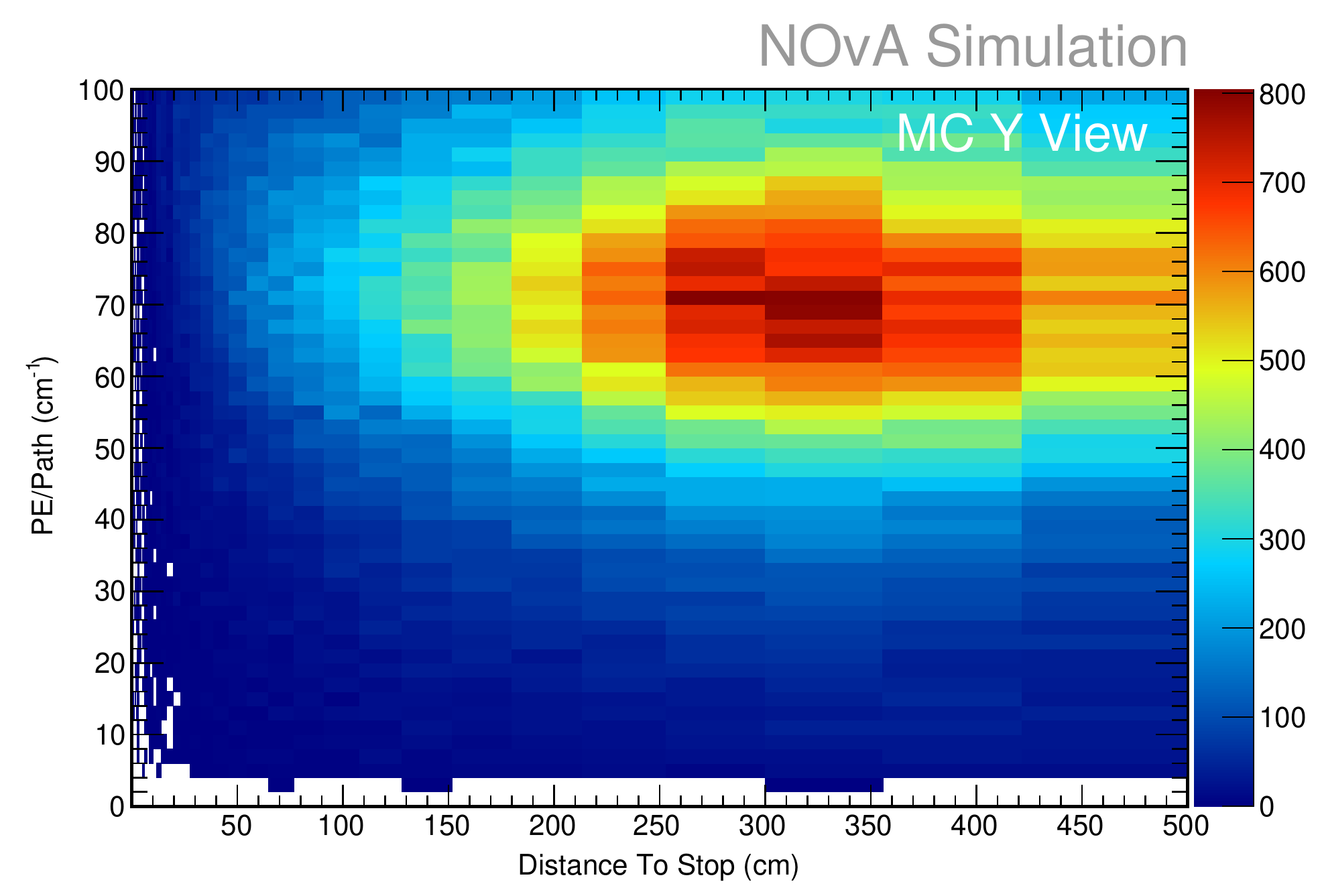}\\
    \includegraphics[width=0.48\textwidth]{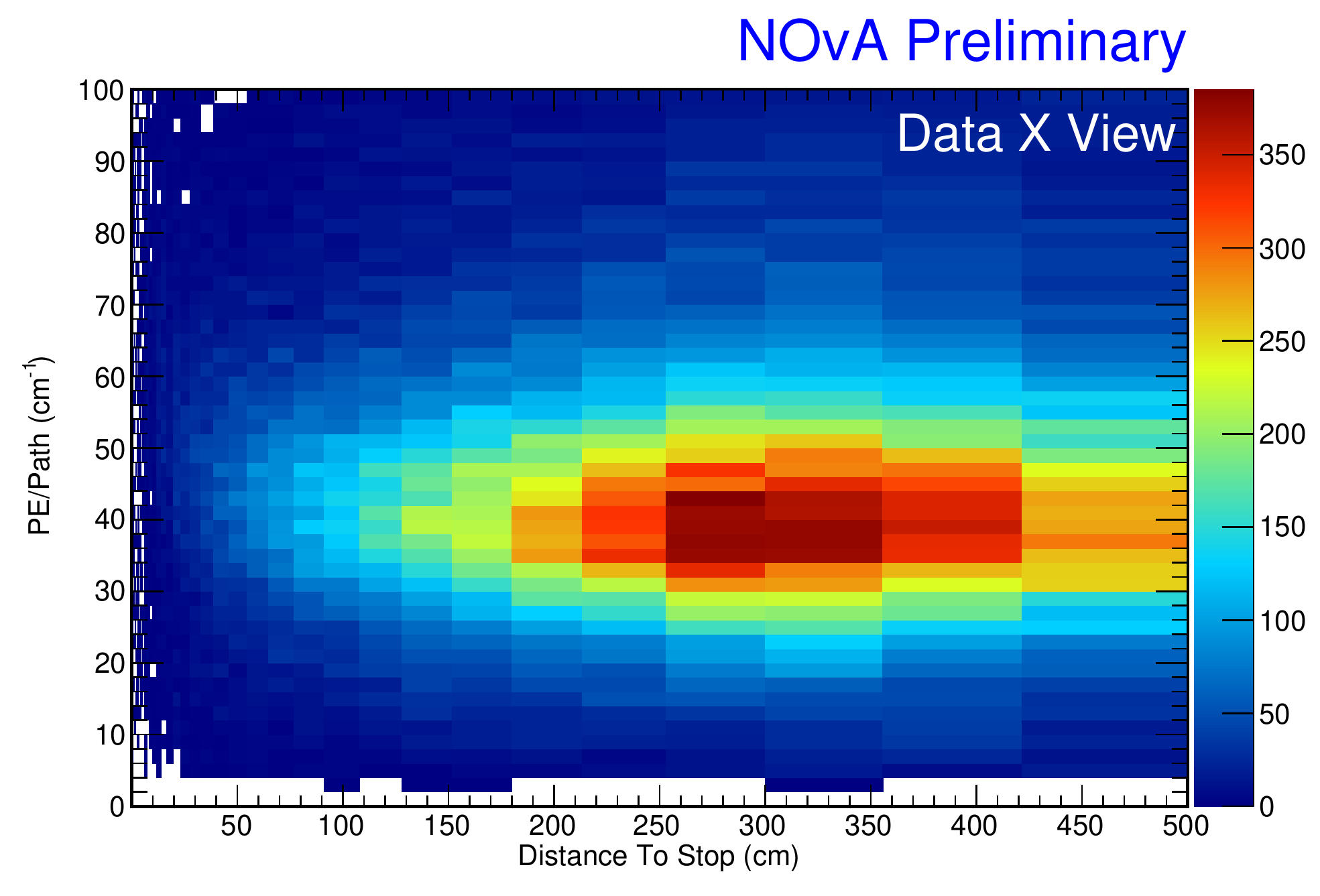}
    \includegraphics[width=0.48\textwidth]{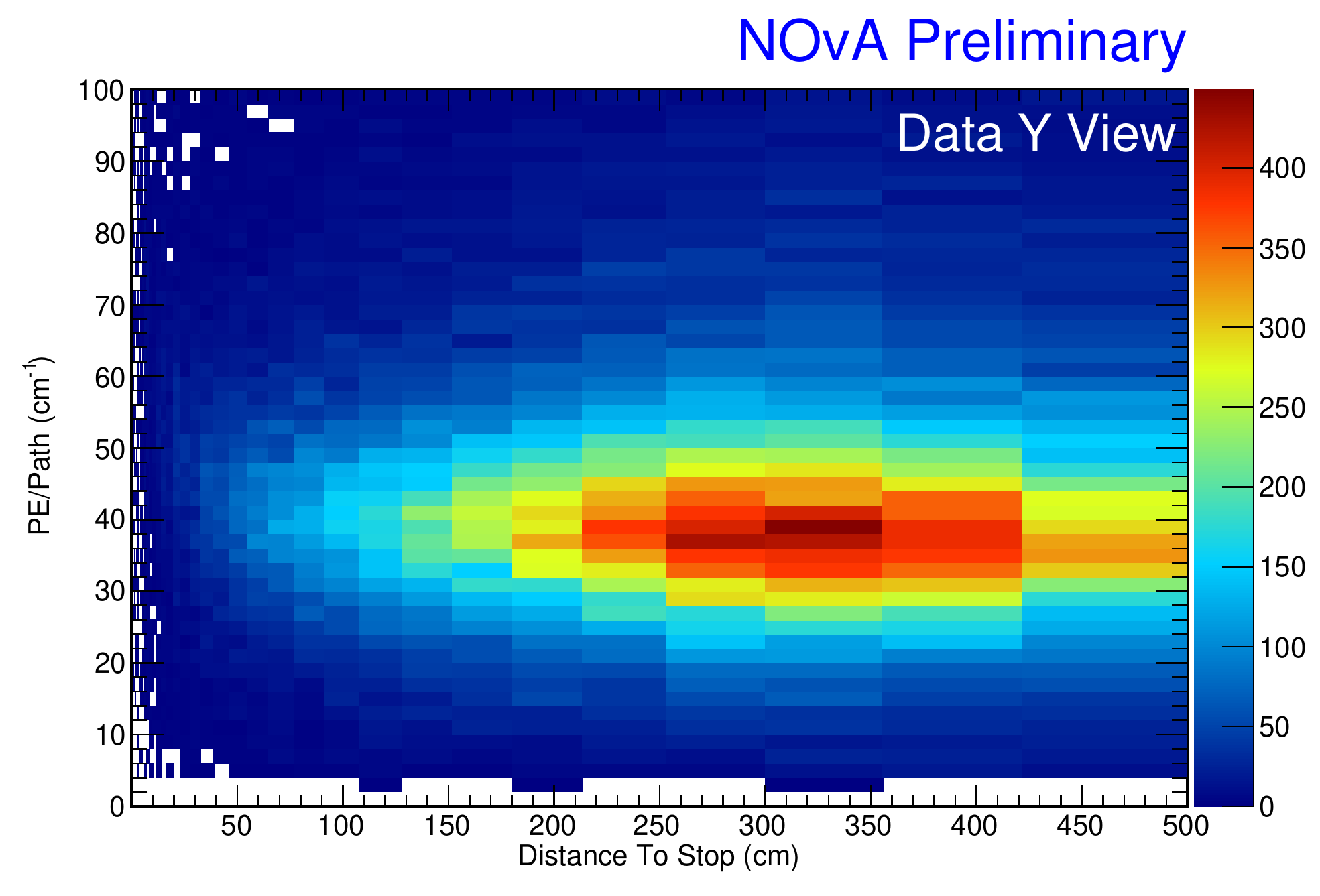}  
    \caption{Selected ND beam muon sample.}
  \end{subfigure}
  \begin{subfigure}{0.48\textwidth}
    \includegraphics[width=0.48\textwidth]{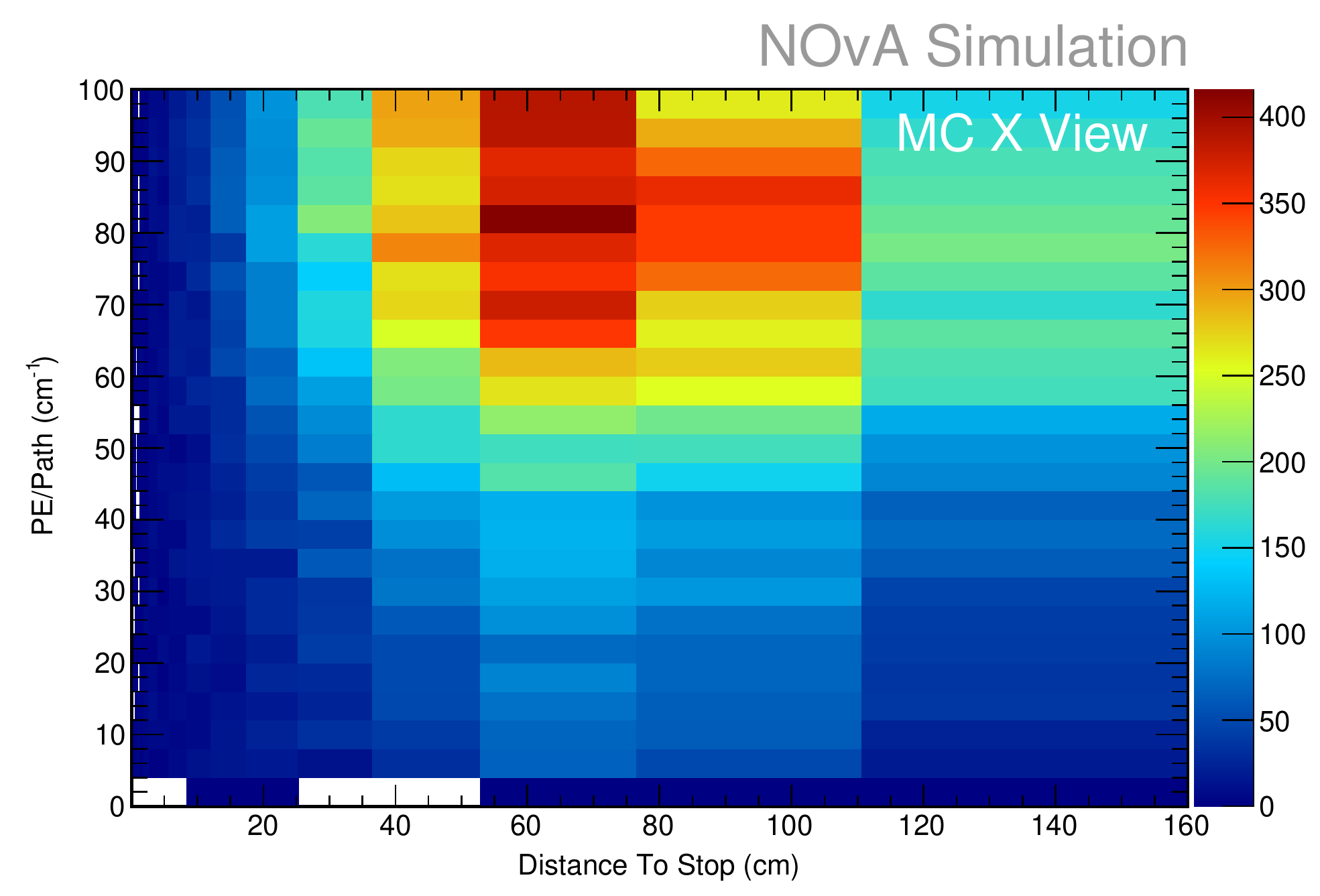}
    \includegraphics[width=0.48\textwidth]{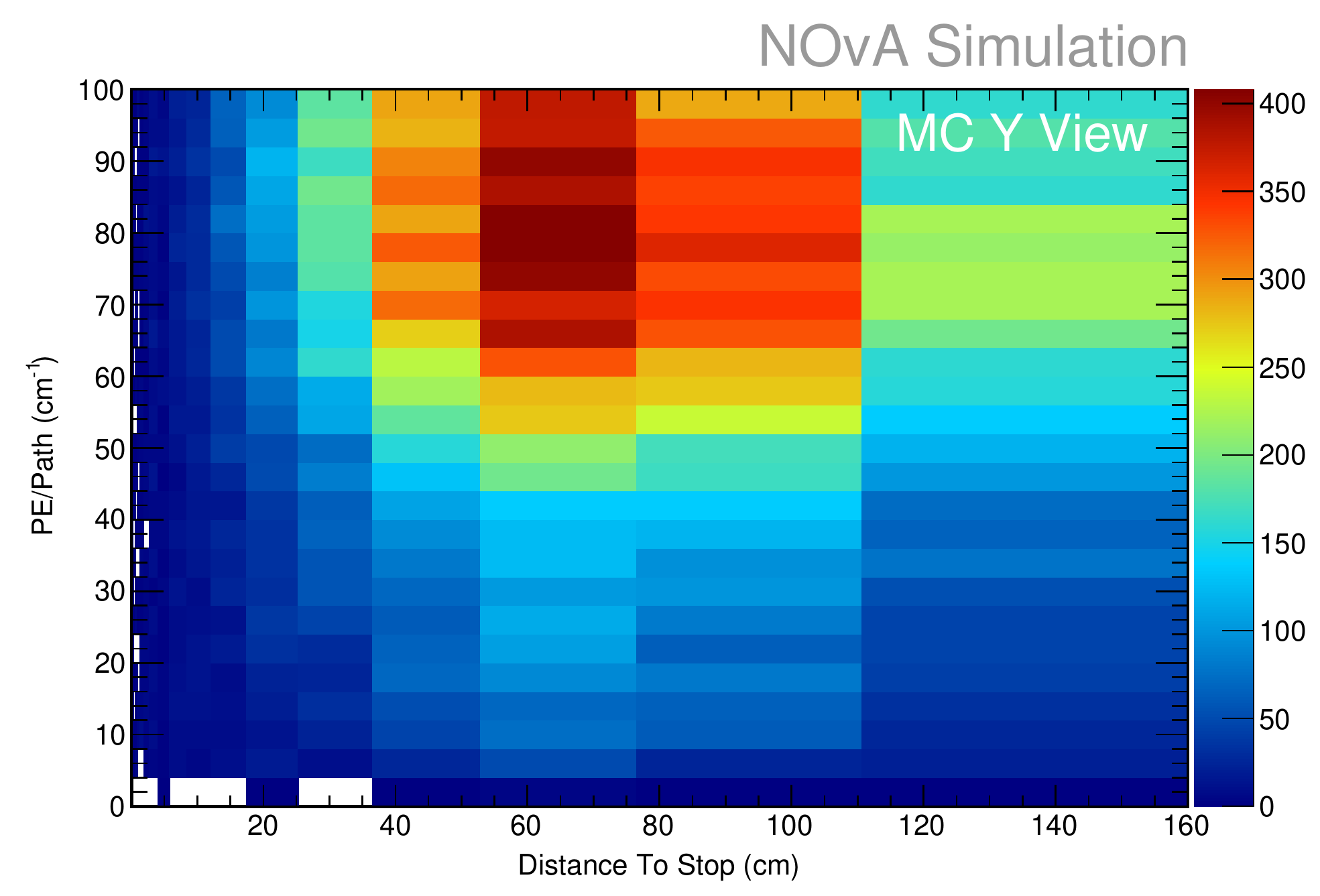}\\
    \includegraphics[width=0.48\textwidth]{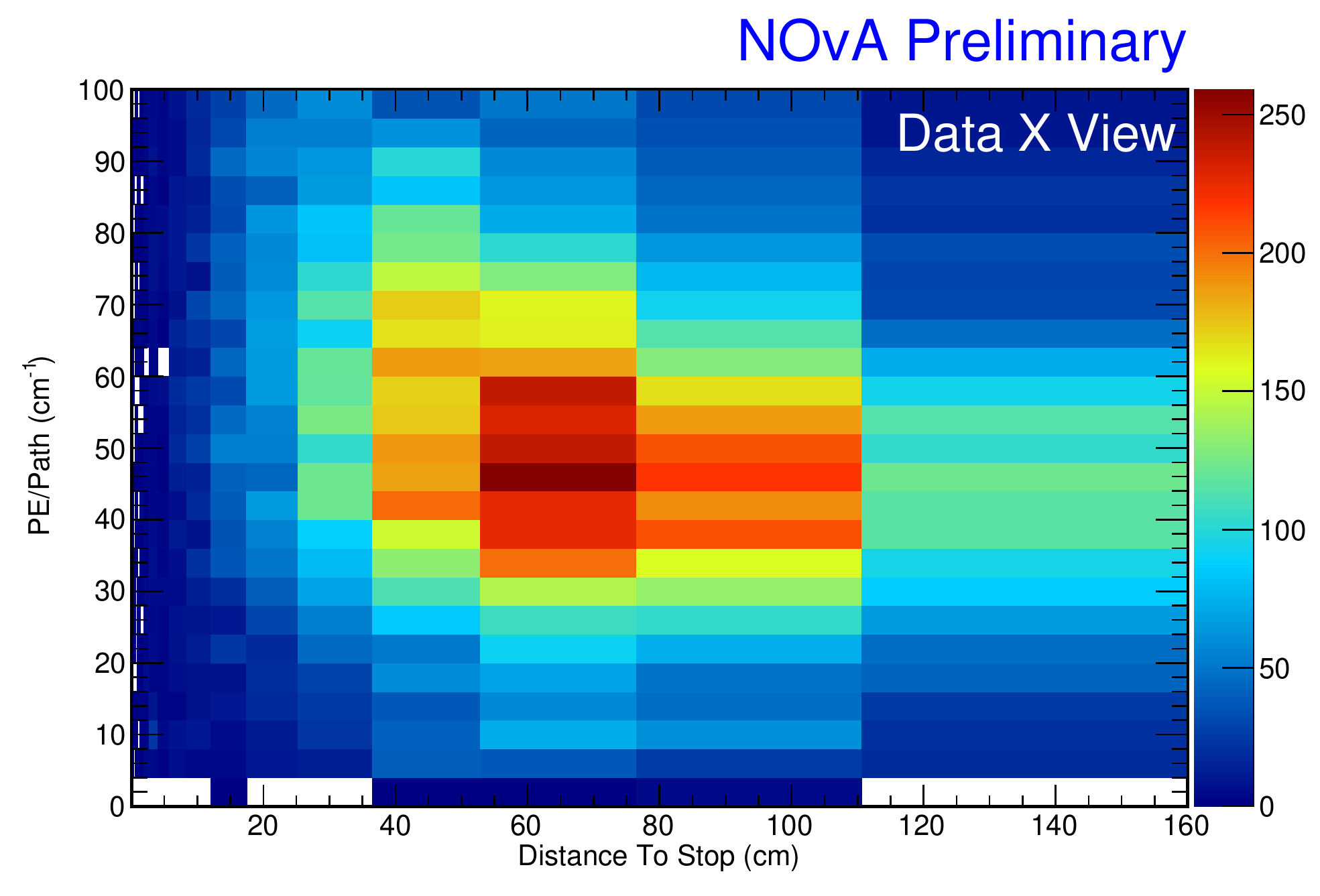}
    \includegraphics[width=0.48\textwidth]{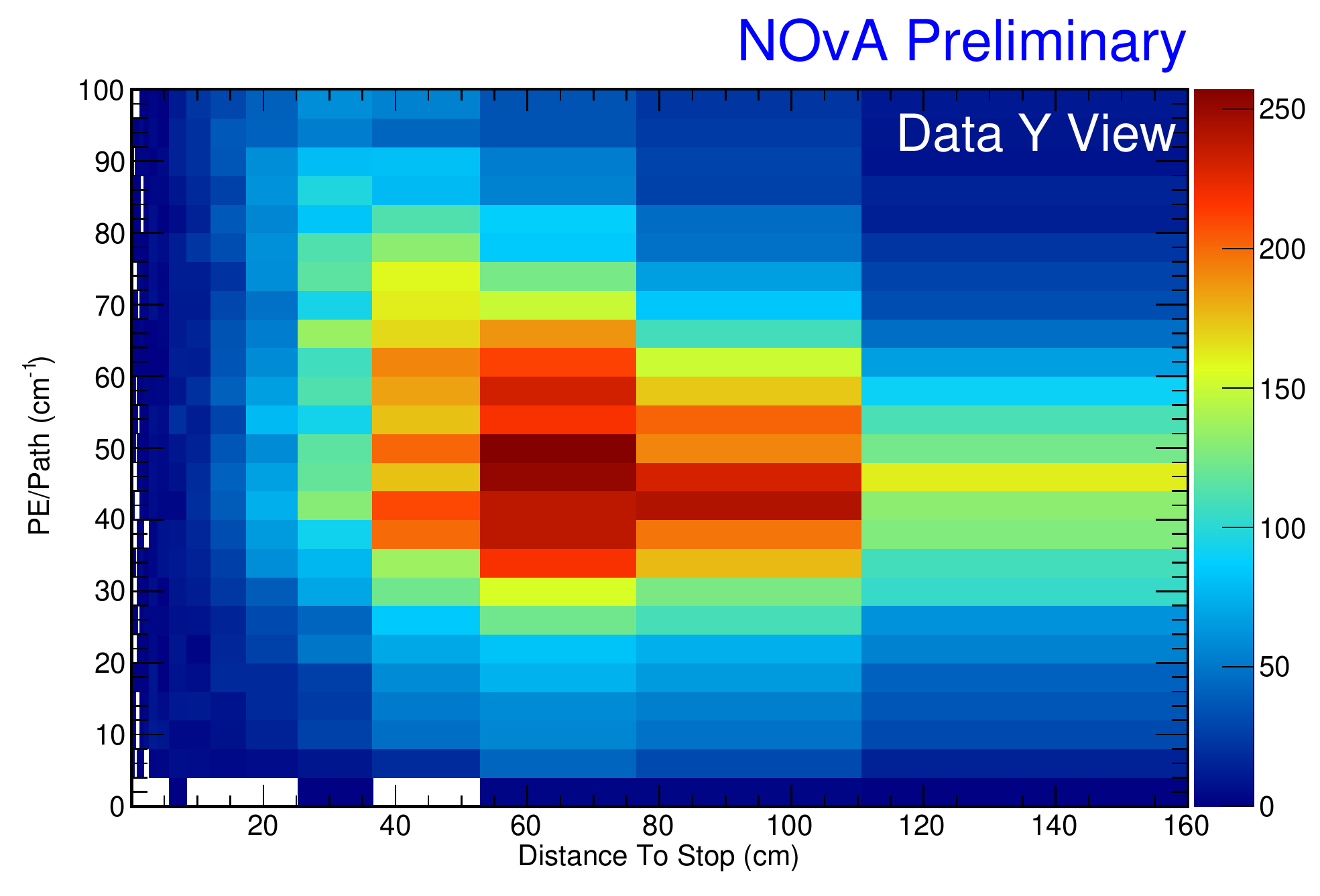}  
    \caption{Selected ND beam proton sample.}
  \end{subfigure}
  
\caption{These spectra are input to the fitting. In each of the sub-figures, the top two plots are MC samples and the bottom two are data samples; plots on the left are from the $xz$ view of the detector and plots on the right are from the $yz$ view of the detector. Binnings are adjusted in different samples for statistical purpose.}
\label{fig:1stprefit}\end{figure}

Since the NOvA detectors have alternating layers, the light model tuning naturally has the view factors tuned separately by splitting the samples into $xz$ and $yz$ views. All the samples are shown in Fig.~\ref{fig:1stprefit}.

\begin{figure}[H]\centering
  \begin{subfigure}{0.35\textwidth}
    \includegraphics[width=\textwidth]{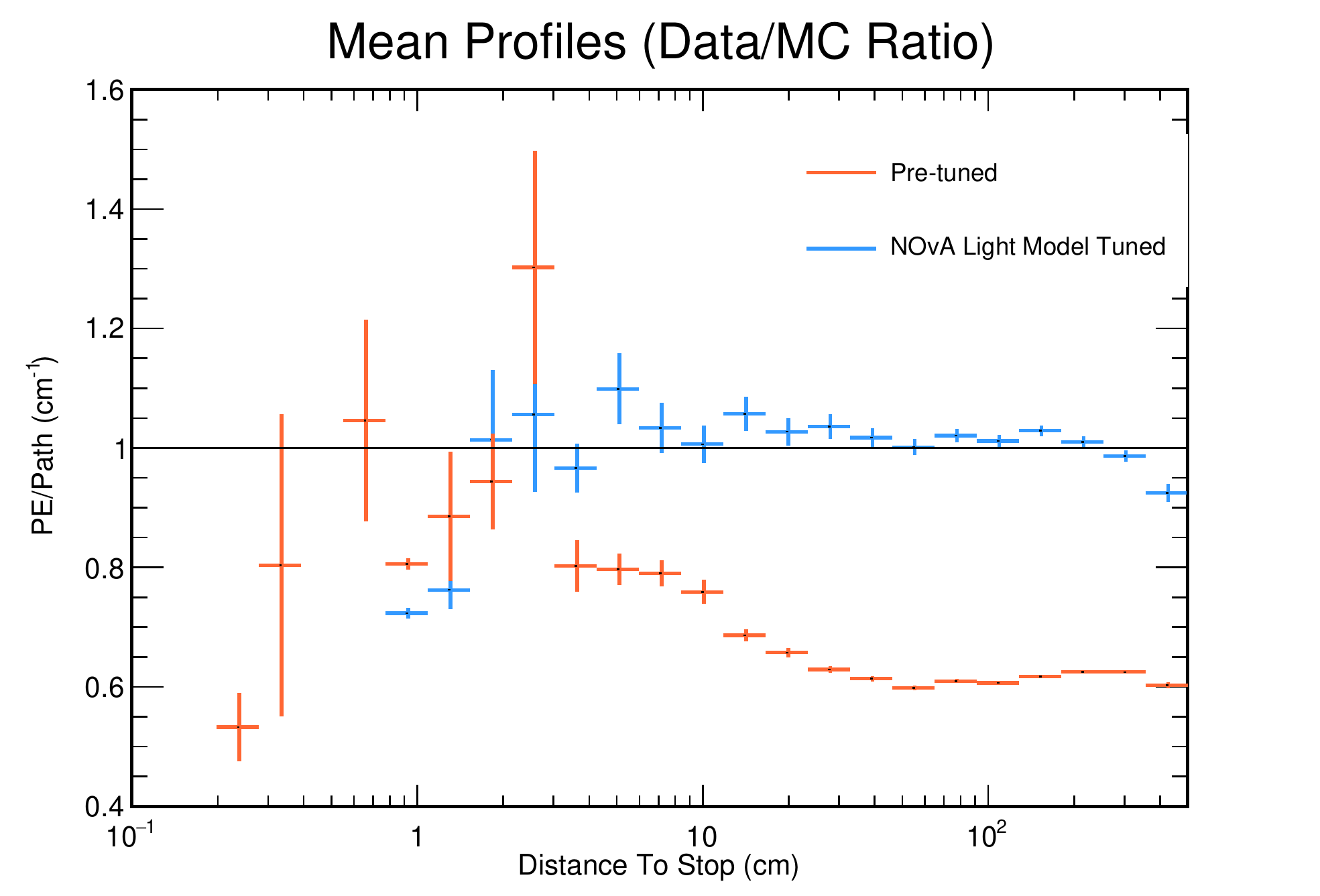}\caption{Selected ND cosmic sample.}
  \end{subfigure}
  \begin{subfigure}{0.35\textwidth}
    \includegraphics[width=\textwidth]{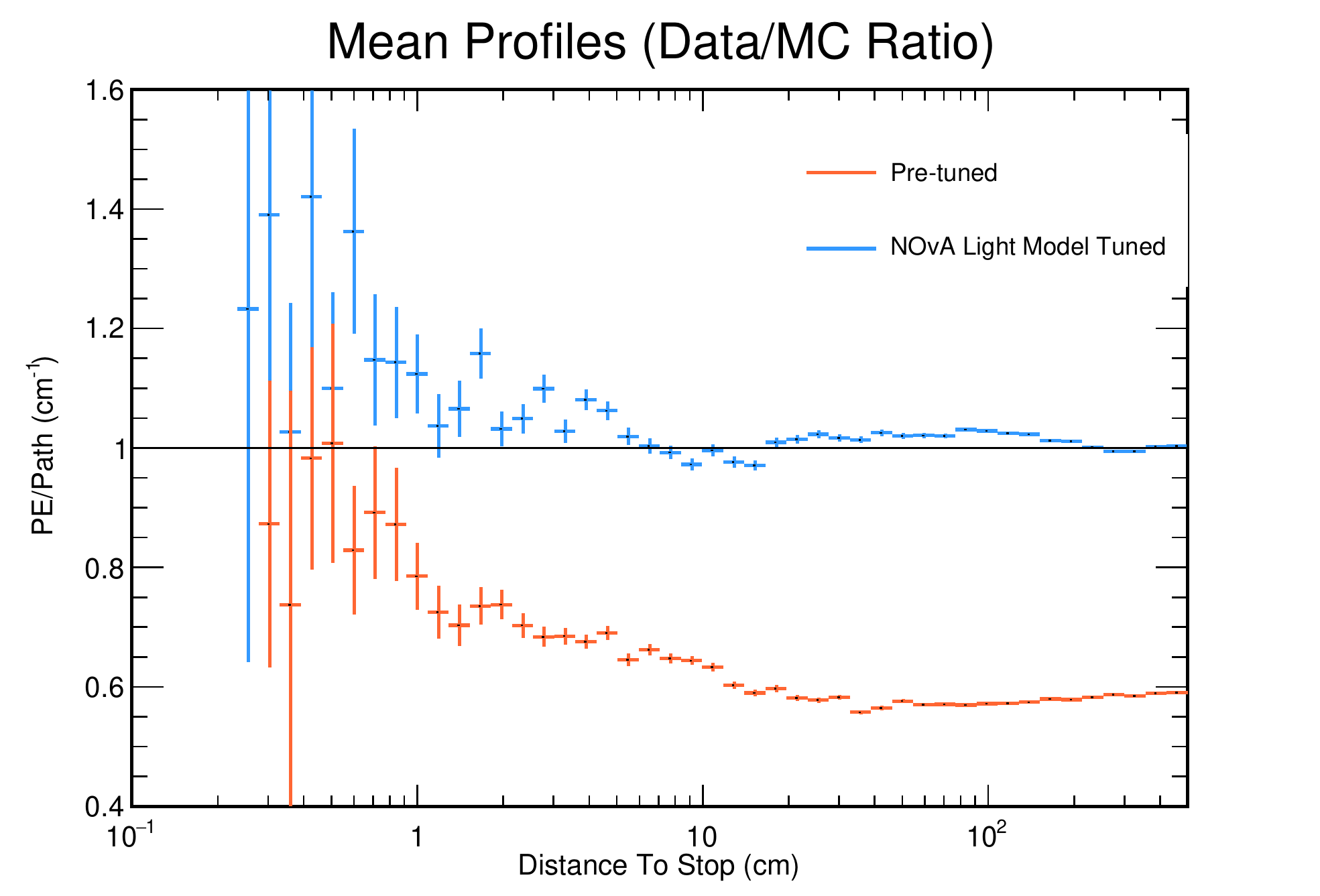}\caption{Selected FD cosmic sample.}
  \end{subfigure}

  \begin{subfigure}{0.35\textwidth}
    \includegraphics[width=\textwidth]{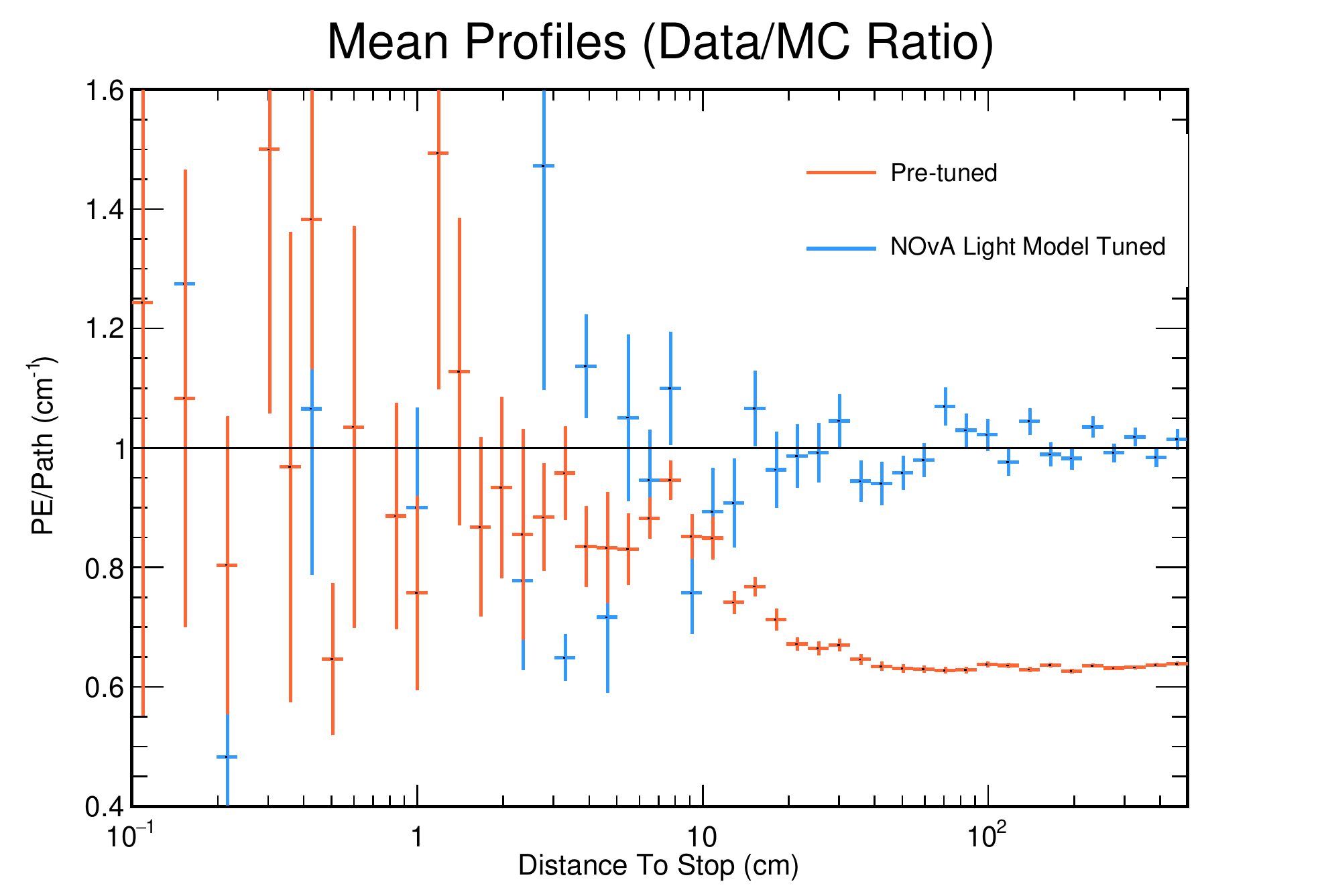}\caption{Selected ND beam muon sample.}
  \end{subfigure}
  \begin{subfigure}{0.35\textwidth}
    \includegraphics[width=\textwidth]{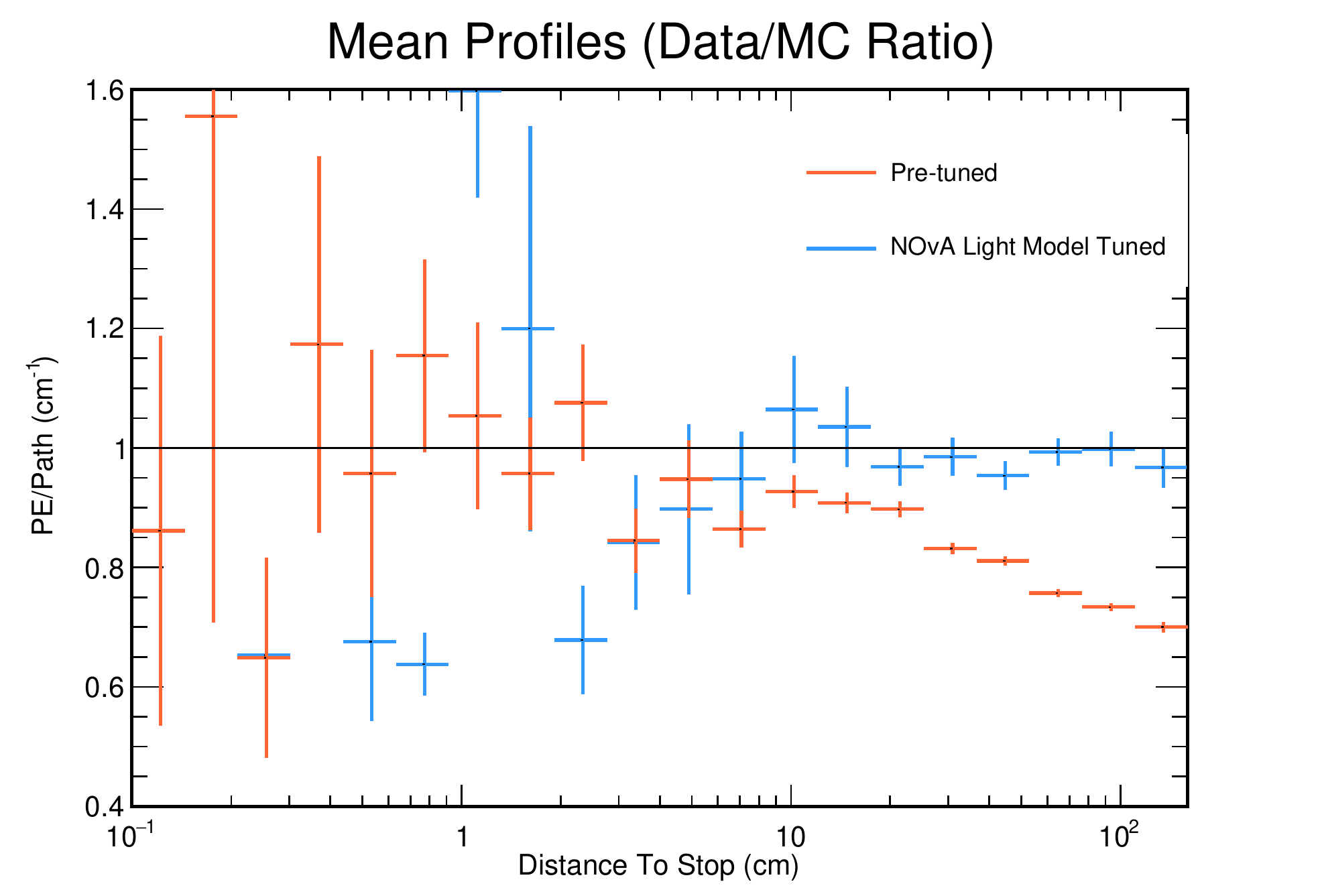}\caption{Selected ND beam proton sample.}
  \end{subfigure}
  
\caption{Data over MC ratios of mean profiles of 2D spectra, as the ones in Fig.~\ref{fig:1stprefit}. Pre-fit vs. post-fit data over MC ratios of detectors' $xz$ views are plotted together for each sample in order to show the improvement on data and MC agreement from applying the new light model parameters.}
\label{fig:xviewproj}\end{figure}

\begin{figure}[H]\centering
  \begin{subfigure}{0.35\textwidth}
    \includegraphics[width=\textwidth]{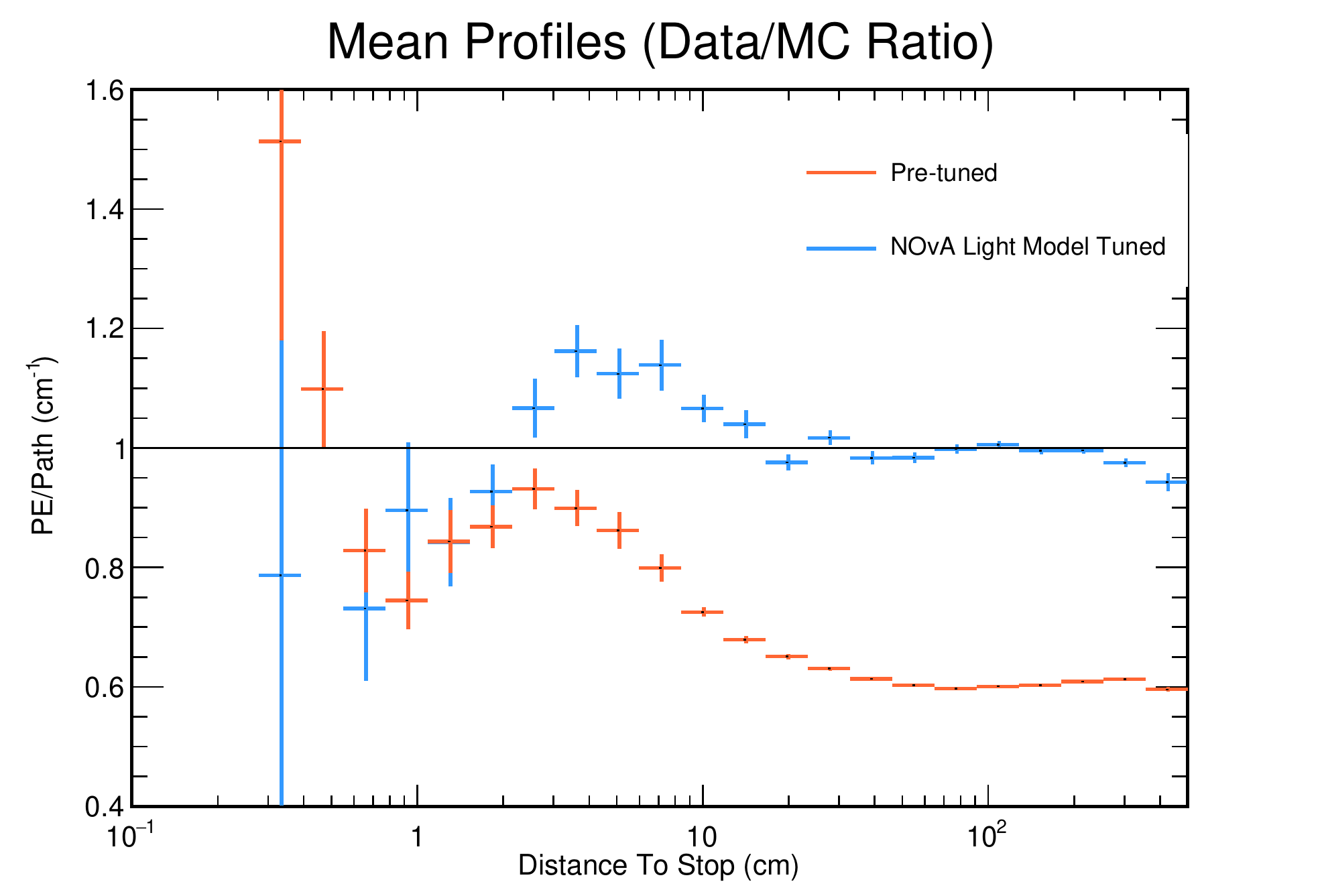}\caption{Selected ND cosmic sample.}
  \end{subfigure}
  \begin{subfigure}{0.35\textwidth}
    \includegraphics[width=\textwidth]{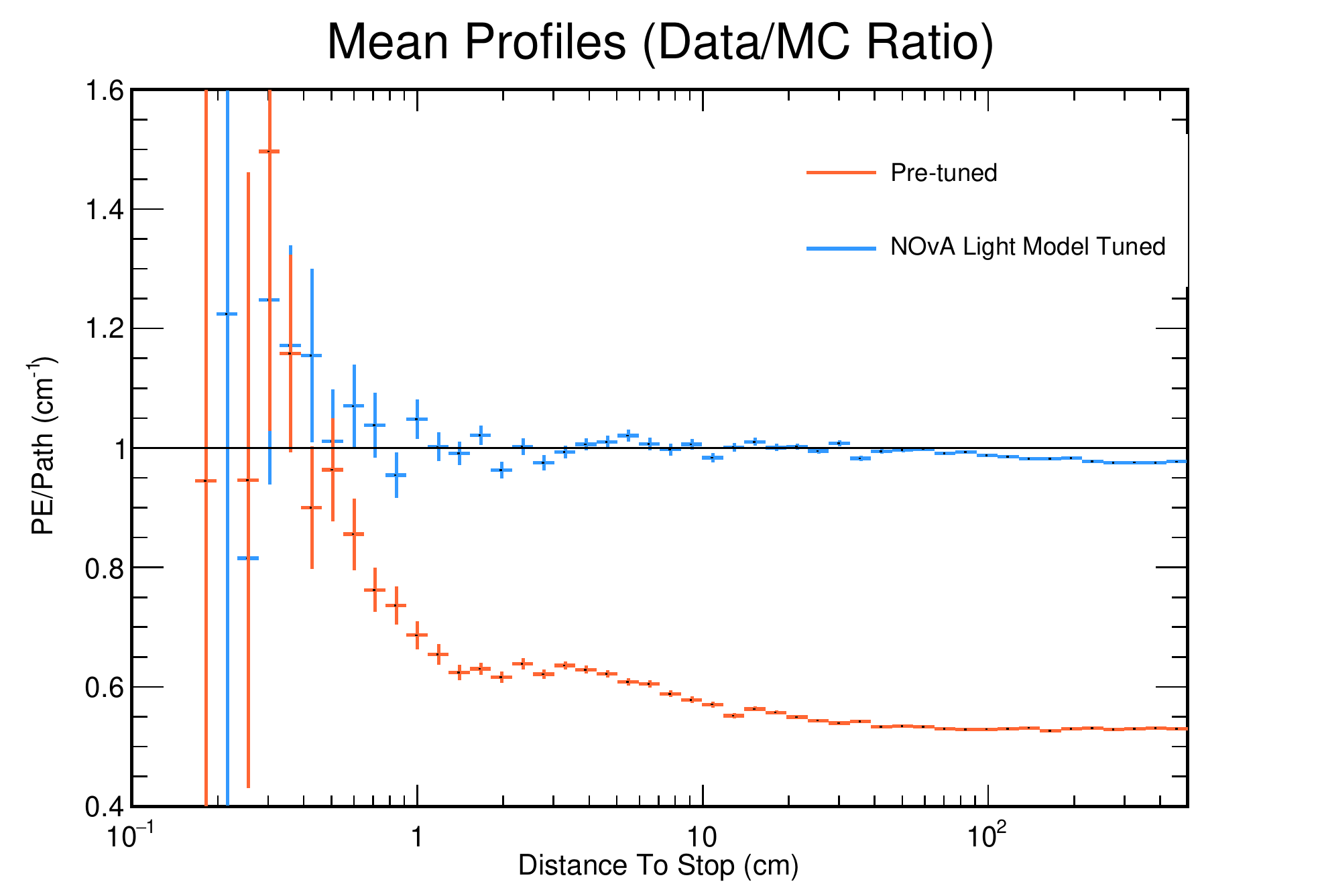}\caption{Selected FD cosmic sample.}
  \end{subfigure}

  \begin{subfigure}{0.35\textwidth}
    \includegraphics[width=\textwidth]{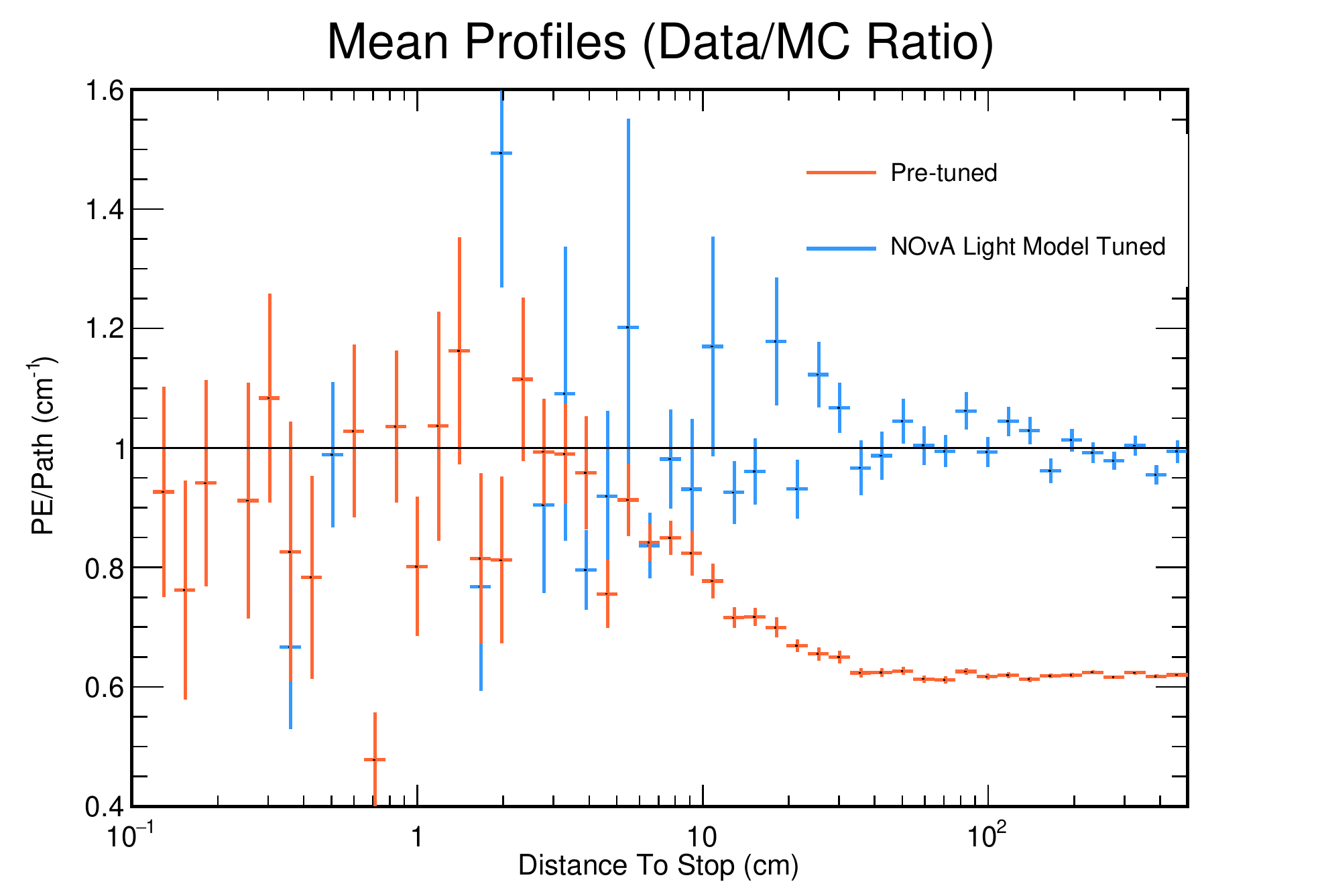}\caption{Selected ND beam muon sample.}
  \end{subfigure}
  \begin{subfigure}{0.35\textwidth}
    \includegraphics[width=\textwidth]{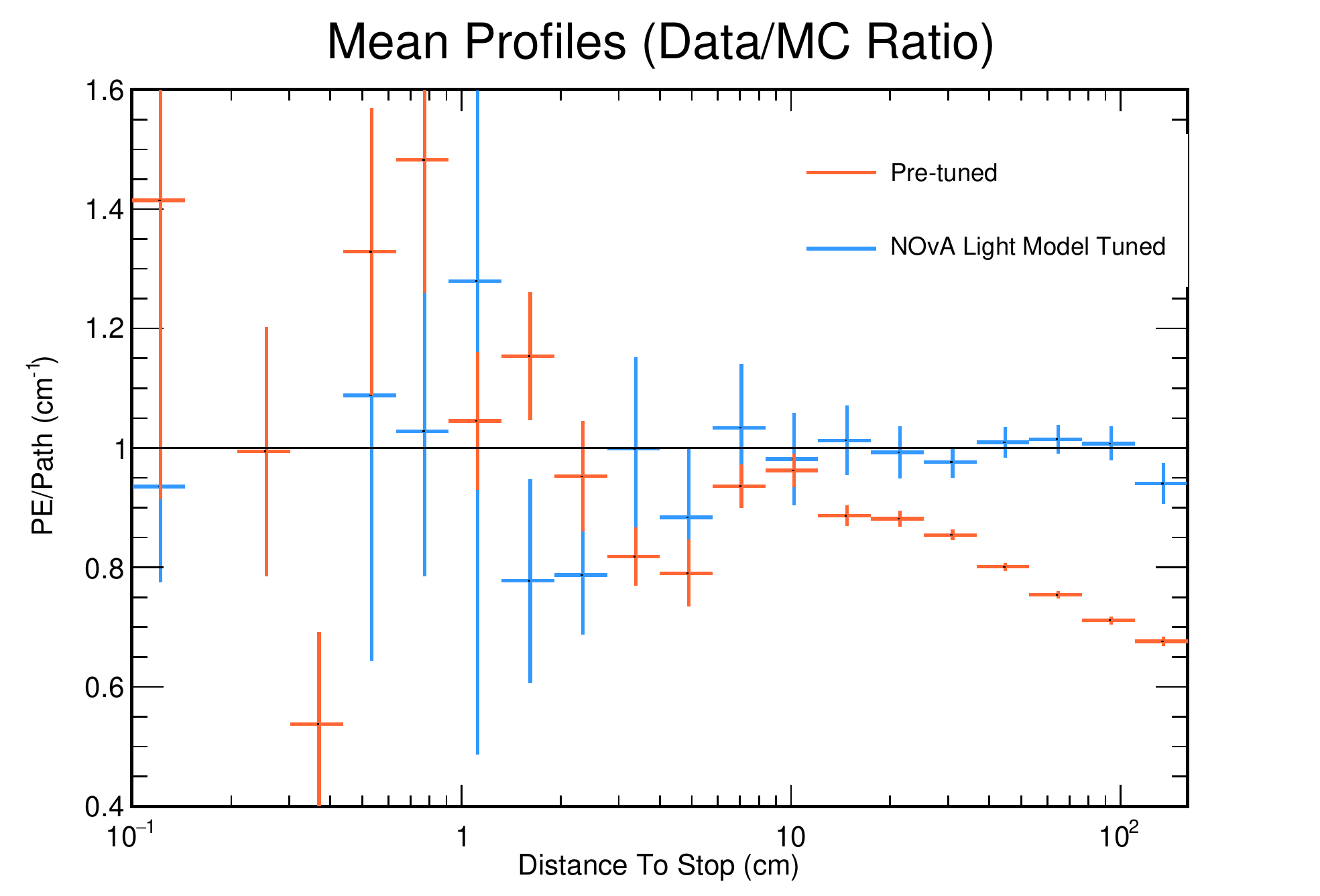}\caption{Selected ND beam proton sample.}
  \end{subfigure}
  
\caption{Data over MC ratios of mean profiles of 2D spectra, as the ones in Fig.~\ref{fig:1stprefit}. Pre-fit vs. post-fit data over MC ratios of detectors' $yz$ views are plotted together for each sample in order to show the improvement on data and MC agreement from applying the new light model parameters.}
\label{fig:yviewproj}\end{figure}

By looking at the mean profiles of the 2D spectra as in Fig.~\ref{fig:1stprefit}, data over MC ratios are compared for before and after applying the new light model parameters extracted in this study. In Fig.~\ref{fig:xviewproj} and Fig.~\ref{fig:yviewproj}, data over MC ratios are flatter and closer to 1 after applying the new light model parameters.

The tuned light model parameters are listed in Table~\ref{2ndresult}. Systematic uncertainties will be evaluated in the future.

\begin{table}[H]\centering
\begin{tabular}{|c|c|c|c|}
\hline
& $Y_{s}$ = 3151.0 & $\epsilon_C$ = 0.471 \\
\hline
ND View Factor &  $F_{xz}$ = 0.58 & $F_{yz}$ = 0.57\\
FD View Factor& $F_{xz}$ = 0.53 & $F_{yz}$ = 0.56\\
\hline
\end{tabular}
\caption{The result of the light model tuning.}\label{2ndresult}
\end{table}

\section*{Acknowledgements}
I am grateful to Department of Energy for supporting the work.

\end{document}

%% file: econfmacros.tex
\def\beq{\begin{equation}}
\def\eeq#1{\label{#1}\end{equation}}
\def\eeqn{\end{equation}}

\def\beqa{\begin{eqnarray}}
\def\eeqa#1{\label{#1}\end{eqnarray}}
\def\eeqan{\end{eqnarray}}

\let\bar=\overbar

\def\Dslash{\not{\hbox{\kern-4pt $D$}}}
\def\dslash{\not{\hbox{\kern-2pt $\del$}}}

\def\msb{{\bar{\ssstyle M \kern -1pt S}}}

